%% file: iclr2027_conference.tex
\documentclass{article} 
\usepackage{iclr2027_conference,times}
\usepackage{graphicx}
\usepackage{subcaption} 
\usepackage{booktabs}
\usepackage{bbm}
\input{math_commands.tex}

\usepackage{hyperref}
\usepackage{url}
\usepackage{enumitem}

\title{A Pitfall in the Evaluation of Gene Perturbation Models}

\author{George Panagopoulos, Saman Omidi, Tsz Pan Tong, \& Jun Pang \\ 
Department of Computer Science\\
University of Luxembourg\\
 2 Av. de l'Universite, 4365 Belval Esch-sur-Alzette, Luxembourg\\
\texttt{\{georgios.panagopoulos,saman.omidi,tszpan.tong,jun.pang\}@uni.lu
}
}

\iclrfinalcopy 
\begin{document}

\maketitle
\lhead{}                              
\renewcommand{\headrulewidth}{0pt}     

\begin{abstract}
Gene perturbation models aim to predict transcriptional responses to unseen genetic interventions, but comparisons rely on heterogeneous evaluation metrics and differentially-expressed (DE) gene subsets. We investigate the consequences of these choices using eight methods across six single-gene perturbation datasets under a unified evaluation protocol. Varying the number of top DE genes reveals substantial instability in Precision-based rankings, with the winning method changing $86.7\%$ of the time, while rankings based on the rest of the metrics remain comparatively stable. Investigating this discrepancy exposes that repeating a single predicted expression profile across cell-level rows, a standard practice for many methods, can inflate significance estimates and produce excessively large predicted significant-DE sets. Specifically we found that $67.1\%$ of tested perturbation--gene pairs pass the significance threshold despite some of them having minuscule effects. 
We derive an inflation factor showing that replicated predictions can amplify the standardized Mann--Whitney statistic even after correction for tied ranks, without adding independent information.
This artifact changes Precision mainly as the number of DE genes increases because for smaller cutoffs, most of the genes are already significant due to the effect-based ordering.
Taking this pitfall into account in our benchmark comparison, we find that a random forest using Gene Ontology features achieves the best overall results. 

\end{abstract}

\section{Introduction}

Large-scale single-cell perturbation screens provide a promising means of studying gene function and identifying potential therapeutic targets. However, experimentally measuring every possible genetic intervention is prohibitively expensive. Gene perturbation models (GPMs) aim to address this limitation by predicting the transcriptional response to previously unseen interventions, typically represented as the change in gene expression relative to non-targeting controls. Recent methods span linear predictors, biologically informed models, graph neural networks, and pretrained foundation models, yet there is still limited agreement on which modeling paradigms perform best.

A major obstacle to reaching consensus is the absence of a common evaluation protocol. Most studies quantify numerical reconstruction, but apply the same families of metrics to substantially different gene set sizes. Error-based metrics such as MSE, MAE, and RMSE are computed on the top-20 observed differentially expressed (DE) genes \citep{roohani2024gears,wenteler2024perteval}, over complete profiles \citep{vcc2025evaluation,adduri2025predicting,scf,lpm}, over the 1,000 most highly expressed genes \citep{ahlmann2025deep}, or over combinations of highly variable and DE genes on different thresholds $N$ 
\citep{wenteler2024perteval,li2026systematic,wei2026benchmarking}. Pearson correlation is similarly reported over all genes \citep{roohani2024gears,scf}, over both all genes and the top-20 DE genes \citep{lpm}, or just in top DE genes \citep{RfgoPaper}. 

Accurate detection of affected genes or perturbations is equally important, but detection metrics also differ in both definition and gene scope. The 2025 Virtual Cell Challenge evaluates recovery of the complete perturbation-specific significant-DE set through its differential-expression score, while the associated Cell-Eval framework includes Precision@\(k\), Overlap@\(k\), and AUPRC \citep{vcc2025evaluation,adduri2025predicting}. Jaccard similarity, Precision@10, and precision--recall analysis are used in \citet{roohani2024gears}, true-positive rate and false-discovery proportion in \citet{ahlmann2025deep}, and AUC and false omission rate in \citet{lpm}. PerturBench combines perturbation-rank metrics with recall of the top-20 DE genes \citep{wu2026perturbench}. \cite{li2026systematic} combine reconstruction metrics computed over all genes and top-20, top-50, and top-100 subsets with top-50 DE gene overlap and direction accuracy. Finally, \cite{radig2026scarchon} use top-20 DE gene overlap and \cite{wei2026benchmarking} combine MSE and PCC-delta with a Common-DE-genes metric, reporting results over all genes and the top-100 DE genes.

These divergent protocols render method rankings difficult to discern. In this work, we treat the evaluation protocol itself as the object of study. We compare eight methods and baselines spanning structured regression, Gene Ontology-derived features, frozen foundation-model representations, graph neural networks, and pretrained perturbation models across six single-gene perturbation datasets. All methods are evaluated using common perturbation-disjoint five-fold splits in unseen perturbation prediction. We systematically vary the number \(N\) of evaluated DE genes and we focus on $@N$ versions of MSE and Pearson for accuracy and Overlap and Precision for detection because they are the most representative. 

We find that Precision-based rankings are particularly sensitive to the number of evaluated DE genes and identify inflated predicted significance as a contributing mechanism. To be specific, ranking based on Precision through different $N$ produces a mean absolute rank displacement of $2.58$ positions and changes the top-ranked method in $86.7\%$ of dataset--fold evaluations. In contrast, MSE, Pearson, and Overlap exhibit mean displacements below one position and winner-change rates no greater than $33.3\%$. Exact dataset-level sign-flip tests identify systematic reordering only for Precision after Benjamini--Hochberg correction. 

Investigating this divergence reveals an artifact in DE testing. The evaluation wrapper repeats a single predicted expression profile across multiple rows to match the ground truth expression matrix, creating a predicted group with no within-condition variation. Mann--Whitney $U$ (MWU) testing against control cells can then assign very small $p$-values to weak predicted shifts. 
We further derive the replication-induced inflation factor of the standardized Mann--Whitney statistic, showing that correction for tied ranks does not account for the dependence among copied predictions. Under the assumptions of our analysis, 100 copies compared with 100 control cells multiply the standardized statistic by approximately $7.62$, despite leaving the predicted expression unchanged and adding no independent information. In practice we find that across $620.9$ million tested perturbation--gene pairs, $67.1\%$ pass the nominal $\mathrm{FDR}<0.05$ threshold, while $8.7\%$ of these significant calls have minuscule predicted effect magnitudes of $|\log_2\mathrm{FC}|<0.1$. 
The resulting predicted significant-DE lists are typically much larger than their ground truth counterparts: only approximately $22\%$ of perturbations have at least $1000$ observed significant genes, and fewer than $1\%$ have at least $4000$. This asymmetry interacts with the metric definitions. Overlap caps its effective cutoff at the observed significant-DE count, so the predicted p-values do not enlarge it. Precision, however, uses the predicted significant-DE count and can therefore continue evaluating inflated predicted lists. Since genes are ordered by absolute log-fold change, smaller $N$ emphasize the strongest predicted effects which are indeed significant, hence the artifact, despite being there, has a limited impact on Precision. However as $N$ increases, weaker effects are admitted by the inflated significance estimates, as proven by our analysis and experiments. 

Taking this pitfall into account, we create a benchmark comparison that aggregates results for small cutoffs of $N$. In this manner, we aim to account for the aforementioned variability between $N$ but reduce the influence of the artifact on Precision. Overall the results agree with the literature that greater architectural complexity does not consistently improve perturbation prediction, but they are not in favor of simple linear models either. A Random Forest with gene ontology features \citep{RfgoPaper} has the best performance overall, followed by the same model with pretrained embeddings from scFoundation \citep{scf}. This shows that informative biological or pretrained representations can be highly effective when combined with comparatively simple predictors.

\section{Related Work}

\paragraph{Known evaluation artifacts in GPMs.}
Recent studies show that GPM evaluations can be distorted by transcriptional shifts shared across perturbations \citep{vinas2025systema}, correlated metrics \citep{radig2026scarchon}, and shared control references \citep{nicol2026spurious}. Other concerns include the inclusion of the directly perturbed gene \citep{RfgoPaper}, sensitivity to perturbation strength \citep{li2026systematic}, and incentives induced by composite scores \citep{vcc2025evaluation}.

Although recent scRNA-seq datasets contain millions of cells, larger cell counts do not necessarily provide more independent biological replicates. Correlated cells from the same biological sample can produce pseudoreplication when treated as independent observations in MWU-based DE testing. Using matched bulk and single-cell datasets, \citet{squair2021confronting} found that pseudobulk methods preserving biological replicate structure improved agreement with bulk DE results and reduced false discoveries. In perturbation prediction, \citet{bhattacharya2026elucidating} show that evaluation metrics respond differently to deterministic and stochastic outputs, while \citet{schafer2026towards} emphasize that cell-count-dependent statistical power complicates comparisons of significance-defined DE gene sets. As datasets and benchmarks proliferate, frameworks such as cell-eval \citep{adduri2025predicting} and DE-gene-recovery metrics such as AUPRC \citep{zhu2025auprc} make careful justification of significance-based evaluation increasingly important. These concerns motivate us to theoretically and quantitatively examine prediction replication used to match observed cell counts in evaluation and examine its consequences for DE gene selection and Precision-based rankings as $N$ increases.

\paragraph{GPM benchmarks.}  Several recent studies have broadened the empirical evaluation of cellular perturbation models. PerturBench \citep{wu2026perturbench} provides standardized datasets, task definitions, model implementations, and metrics for genetic and chemical perturbations, including covariate-transfer and combinatorial-prediction settings. Foundation-model-focused evaluations include PertEval-scFM \citep{wenteler2024perteval}, which examines whether frozen single-cell foundation-model representations improve zero-shot genetic perturbation prediction, and \cite{RfgoPaper}, who compare scGPT and scFoundation with train-mean and biologically informed classical baselines across four Perturb-seq datasets. The latter study finds that simple baselines can outperform foundation models and identifies low perturbation-specific variance as a limitation of commonly used benchmark datasets.  In a broader biomedical setting, \cite{el2026transcriptomic} introduce a transcriptomic foundation-model benchmark comprising 35 tasks across eight immunological and inflammatory diseases, spanning target discovery, preclinical translation, treatment-response prediction, and patient stratification; perturbation prediction is one component of this wider drug-development benchmark.  Broader comparisons of gene perturbation models have also rapidly expanded in scale. 
\cite{wei2026benchmarking} compare 27 methods across 29 datasets under perturbation- and cellular-context generalization settings while \cite{li2026systematic} conduct a task-spanning comparison of 13 methods on 25 datasets using 24 metrics covering expression accuracy, differential-expression recovery, and distributional similarity. 
\cite{radig2026scarchon} evaluate the transfer of non-genetic perturbation responses across cell types and individuals rather than unseen genetic interventions.

\section{Experimental Design}


\subsection{Evaluation Metrics}
\label{sec:metrics}


We evaluate model performance with metrics computed on increasingly large sets of differentially expressed (DE) genes and over the complete gene set. The finite-\(N\) analysis uses $N\in\{50,100,200,1000,2000,4000\}$ and includes two metric families $\text{Detection}=\{\text{Overlap@}N,\text{Precision@}N\}$, and numerical accuracy $\text{Accuracy}=\{\text{MSE@}N,\text{Pearson@}N\}$ as explained below.

We use \textbf{cell-eval} \citep{adduri2025predicting} to compute differential-expression statistics from observed and predicted profiles. Each model receives the name of the perturbation as input, and its output is represented as a vector $\widehat{\Delta}\in\mathbb{R}^{|G|}$ corresponding to the predicted effect. For methods producing a single profile per perturbation, this vector is repeated across the prediction rows into an effect matrix $\widehat{\mathbf{\Delta}} \in\mathbb{R}^{|C|\times|G|}$,
where $C$ is the set of observed cells receiving that perturbation, $G$ is the set of expressed genes in the profile. The observed effect matrix $\mathbf{\Delta}$ has the same dimensions so that they can be compared.

The average control expression is added to each row of $\widehat{\mathbf{\Delta}}$ to derive the predicted post-perturbation expression matrix:
\[
\overline{x}_{\mathrm{ctrl}}
=
\frac{1}{|C_{\mathrm{ctrl}}|}
\sum_{j\in C_{\mathrm{ctrl}}}x_j^{\mathrm{ctrl}},
\qquad
\widehat{\mathbf{X}}
=
\widehat{\mathbf{\Delta}}
+
\mathbf{1}_{|C|}\overline{x}_{\mathrm{ctrl}}^{\top},
\]
where $C_{\mathrm{ctrl}}$ is the set of non-targeting control cells and is not necessarily equal to $|C|$, and $x_j^{\mathrm{ctrl}}\in\mathbb{R}^{|G|}$ is the expression profile of control cell $j$ and $\mathbf{X}\in\mathbb{R}^{|C|\times|G|}$ be the observed post-perturbation expression matrix.

For each perturbation and each tested gene, DE is assessed separately by comparing
(i) expression across observed perturbed cells with expression across non-targeting control cells, and
(ii) expression across predicted rows with expression across the same control cells. Each comparison uses the Mann--Whitney $U$ test \citep{mann1947test}, a nonparametric test based on the ranks of expression values pooled across the two groups. It tests the null hypothesis that the groups have the same underlying expression distribution, without assuming normality. Thus, each gene receives separate observed and predicted $p$-values:
\[
\begin{aligned}
p_g^{\mathrm{obs}}
&=
\mathrm{MWU}
\left(
(X_{ig})_{i\in C},
(X_{jg}^{\mathrm{ctrl}})_{j\in C_{\mathrm{ctrl}}}
\right),\\
p_g^{\mathrm{pred}}
&=
\mathrm{MWU}
\left(
(\widehat{X}_{ig})_{i\in C},
(X_{jg}^{\mathrm{ctrl}})_{j\in C_{\mathrm{ctrl}}}
\right),
\end{aligned}
\]
where $\mathrm{MWU}$ returns the $p$-value from the test. Predicted expression is not tested directly against observed perturbed expression. Within each perturbation, the $p$-values are adjusted across tested genes using the Benjamini--Hochberg procedure \citep{benjamini1995controlling}, separately for the observed and predicted comparisons. Genes with adjusted $p<0.05$, denoted $\mathrm{FDR}<0.05$, are retained as significant DE genes. We exclude the directly perturbed gene following the observation of \citet{RfgoPaper} and rank the remaining genes by decreasing absolute log-fold change relative to controls, using observed fold changes for the observed list and predicted fold changes for the predicted list: 
\[
\begin{aligned}
R &= \operatorname{sort}_{|\mathrm{LFC}^{\mathrm{obs}}_g|\downarrow}
\left(\left\{g : p_g^{\mathrm{obs}}<0.05,\; g\neq g_\star\right\}\right),\\
P &= \operatorname{sort}_{|\mathrm{LFC}^{\mathrm{pred}}_g|\downarrow}
\left(\left\{g : p_g^{\mathrm{pred}}<0.05,\; g\neq g_\star\right\}\right),
\end{aligned}
\]
where $g_\star$ is the directly perturbed gene and sorting is by
decreasing absolute observed or predicted log-fold change relative
to controls, respectively. Thus, $p$-values determine list membership,
while fold-change magnitudes determine ordering.
Let
\[
R=(r_1,\ldots,r_{|R|})
\qquad\text{and}\qquad
P=(p_1,\ldots,p_{|P|})
\]
denote the resulting ranked lists of observed and predicted significant DE genes, respectively. Reconstruction metrics are evaluated on the strongest available observed DE genes,
\[
S_N=R_{1:K_R},
\qquad
K_R=\min(N,|R|).
\]

For each perturbation, we compute
$\text{MSE@}N \text{ and } \text{Pearson@}N$
using the signed observed and predicted perturbation effects averaged across cells and prediction rows, respectively. For each gene $g\in S_N$, these are defined as
\[
\Delta_g
=
\frac{1}{|C|}
\sum_{i\in C}(\mathbf{\Delta})_{ig},
\qquad
\widehat{\Delta}_g
=
\frac{1}{|C|}
\sum_{i\in C}(\widehat{\mathbf{\Delta}})_{ig}.
\]
The metrics are then computed across the selected observed significant DE genes:
\[
\mathrm{MSE@}N
=
\frac{1}{K_R}
\sum_{g\in S_N}
\left(\widehat{\Delta}_g-\Delta_g\right)^2,
\]
and
\[
\mathrm{Pearson@}N
=
\operatorname{corr}
\left(
(\widehat{\Delta}_g)_{g\in S_N},
(\Delta_g)_{g\in S_N}
\right),
\]
where $\operatorname{corr}$ denotes Pearson correlation across genes and $K_R=|S_N|=\min(N,|R|)$.


$\text{Overlap@}N \text{ and } \text{Precision@}N$  compare the identities of the observed and predicted significant DE genes, irrespective of effect direction. Overlap uses an effective cutoff determined by the number of observed significant genes,
\[
\mathrm{Overlap@}N
=
\frac{
\left|R_{1:K_R}\cap P_{1:K_R}\right|
}{
K_R
},
\qquad
K_R=\min(N,|R|),
\]
whereas Precision uses a cutoff determined by the number of predicted significant genes,
\[
\mathrm{Precision@}N
=
\frac{
\left|R_{1:K_P}\cap P_{1:K_P}\right|
}{
K_P
},
\qquad
K_P=\min(N,|P|).
\]

Thus, Precision@$N$ uses a model-dependent effective cutoff because $|P|$ varies across models and perturbations, whereas Overlap@$N$ uses an observed-data-dependent cutoff that is shared across models for a given perturbation. When both $R$ and $P$ contain at least $N$ genes, $K_R=K_P=N$ and the two scores are identical. Perturbations with $K_P=0$ are excluded from Precision@$N$, while those with $K_R=0$ are excluded from Overlap@$N$. For brevity, we omit the @$N$ suffix from metric names throughout the remainder of the text. 




\subsection{Methods}

\begin{table}[!ht]
    \centering
    \caption{
    Overview of the evaluated methods. ``Prior'' denotes an explicit
    structured biological prior, while ``Pretrained'' indicates that the
    model or representation was learned from external data before training
    the benchmark-specific predictor.
    }
    \label{tab:method_overview}
        \begin{tabular}{llllc}
            \toprule
            \textbf{Method} & \textbf{Representation} & \textbf{Predictor} & \textbf{Prior} & \textbf{Pretrained} \\
            \midrule
            Bilinear & Low-dimensional factors & Bilinear regression & None & No \\
            RF-GO & GO-derived gene features & Random forest & GO & No \\
            Ridge-scf & scFoundation embeddings & Ridge regression & None explicit & Yes \\
            RF-scf & scFoundation embeddings & Random forest & None explicit & Yes \\
            GEARS & GO node embeddings & GNN & GO & No \\
            LPM & Perturbation embeddings & NN & None explicit & No \\
            Average Control & Mean control profile & - & None & No \\
            Average Train & Mean training profile & - & None & No \\
            \bottomrule
        \end{tabular}%
\end{table}

We study the prediction of transcriptional responses to previously unseen genetic perturbations. 
We benchmarked a diverse set of perturbation-response prediction methods, including the bilinear model of \cite{ahlmann2025deep}, a random forest with Gene Ontology (GO) features (RF-GO) \citep{RfgoPaper}, ridge regression using scFoundation representations (Ridge-scf) \citep{scf}, random forest using pretrained scFoundation representations (RF-scf) \citep{scf}, the graph neural network GEARS that also relies on GO \citep{roohani2024gears}, and the Large Perturbation Model (LPM) \citep{lpm}. We additionally include Average control (AVG Ctrl) and Average train (AVG Train) baselines \citep{vinas2025systema} which are computed throughout the whole sets, not per batch. All methods are shown in Table \ref{tab:method_overview}.
All methods are evaluated using the same perturbation-disjoint five-fold cross-validation splits, such that each test perturbation is absent from the corresponding training fold. Hyperparameters are fixed to the default configurations of the corresponding implementations, and no hyperparameter optimization is performed. A brief explanation of each method along with implementation details is given in App. \ref{sec:app_methods}.

\subsection{Datasets}


\begin{table}[t]
    \centering
    \caption{Dataset characteristics. Mean and median refer to cells
    per perturbation; perturbed and control counts include all
    perturbed and non-targeting control cells, respectively.
    ``Genes'' denotes the number of retained expression features.}
    \label{tab:datasets}
    \label{tab:datasets_full}
    \setlength{\tabcolsep}{3pt}
    \resizebox{\linewidth}{!}{%
    \begin{tabular}{@{}lrrrrrr@{}}
        \toprule
        & &
        \multicolumn{2}{c}{Cells per perturbation} &
        \multicolumn{2}{c}{Cell counts} & \\
        \cmidrule(lr){3-4}
        \cmidrule(lr){5-6}
        Dataset & Perturbations
        & Mean & Median & Perturbed & Control & Genes \\
        \midrule
        Jurkat~\citep{nadig2025transcriptome}
        & 2{,}393 & 104.9 & 83.0
        & 250{,}943 & 12{,}013 & 8{,}882 \\

        RPE1~\citep{Replogle2022}
        & 2{,}393 & 98.8 & 72.0
        & 236{,}429 & 11{,}485 & 8{,}749 \\

        HepG2~\citep{nadig2025transcriptome}
        & 2{,}393 & 58.7 & 45.0
        & 140{,}497 & 4{,}976 & 9{,}624 \\

        K562~\citep{Replogle2022}
        & 2{,}057 & 145.7 & 121.0
        & 299{,}694 & 10{,}691 & 8{,}563 \\

        VCC~\citep{youngblut2025scbasecount}
        & 150 & 1{,}220.6 & 1{,}045.0
        & 183{,}097 & 38{,}176 & 18{,}080 \\

        Adamson~\citep{adamson2016multiplexed}
        & 86 & 515.6 & 504.5
        & 44{,}340 & 24{,}263 & 5{,}060 \\
        \bottomrule
    \end{tabular}%
    }
\end{table}

We use six single-gene perturbation datasets spanning different
perturbation scales, number of cells, and experimental settings
based on single-cell RNA-sequencing experiments using
CRISPR interference (CRISPRi). As summarized in
Table~\ref{tab:datasets}, the datasets contain between 86 and
2{,}393 perturbations, with an average of approximately 59 to
1{,}221 cells per perturbation. Median cell counts are below
the corresponding means in every dataset, consistent with
unequal sampling across perturbations. Control cells account
for approximately 3--5\% of all cells in Jurkat, RPE1, HepG2,
and K562, compared with 17.3\% in VCC and 35.4\% in Adamson. The preprocessing pipeline along with a comparison between datasets are included in App. \ref{sec:preprocessing}.

\section{Results}

\begin{figure}[h!]
    \centering

    \begin{subfigure}[t]{0.51\linewidth}
        \vspace{0pt}
        \centering
        \includegraphics[width=\linewidth]{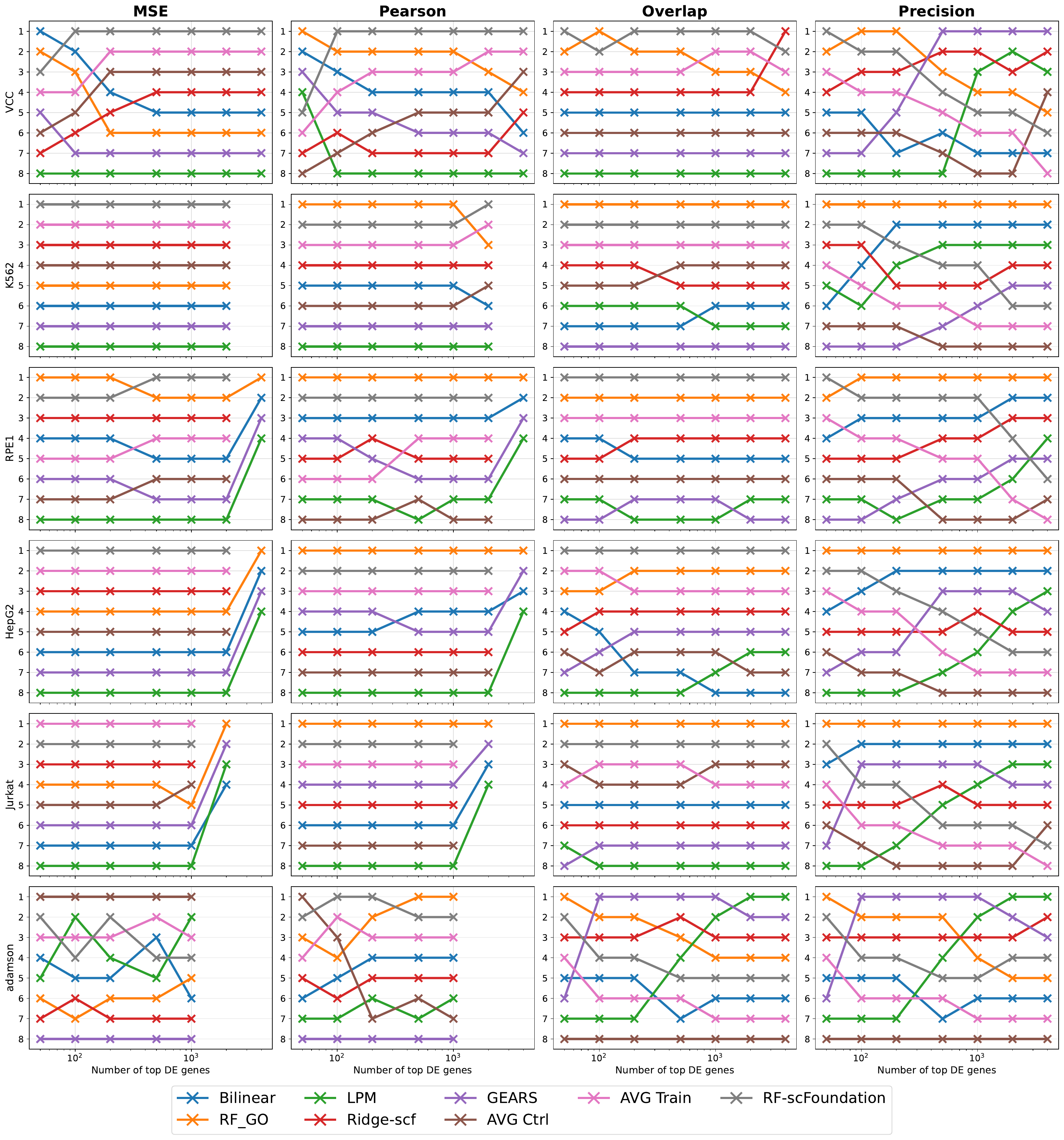}
        \caption{}
        \label{fig:rank_cross}
    \end{subfigure}
    \hfill
    \begin{minipage}[t]{0.48\linewidth}
        \vspace{0pt}
        \centering

        \begin{subfigure}[t]{\linewidth}
            \centering
            \includegraphics[width=0.62\linewidth]
            {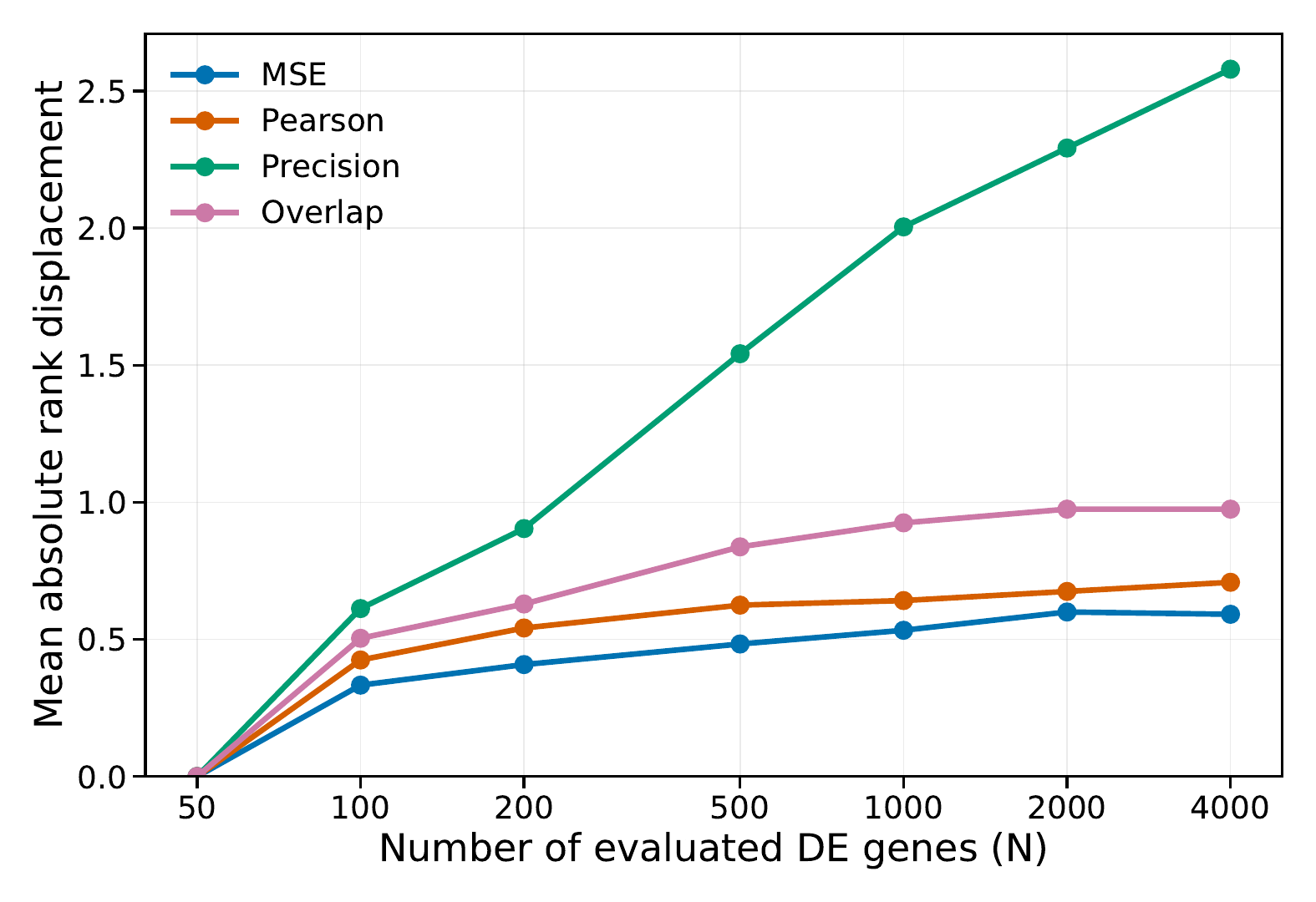}
            \caption{}
            \label{fig:rank_displacement}
        \end{subfigure}

        \vspace{1em}

        \begin{subfigure}[t]{\linewidth}
            \centering
            \includegraphics[width=0.62\linewidth]
            {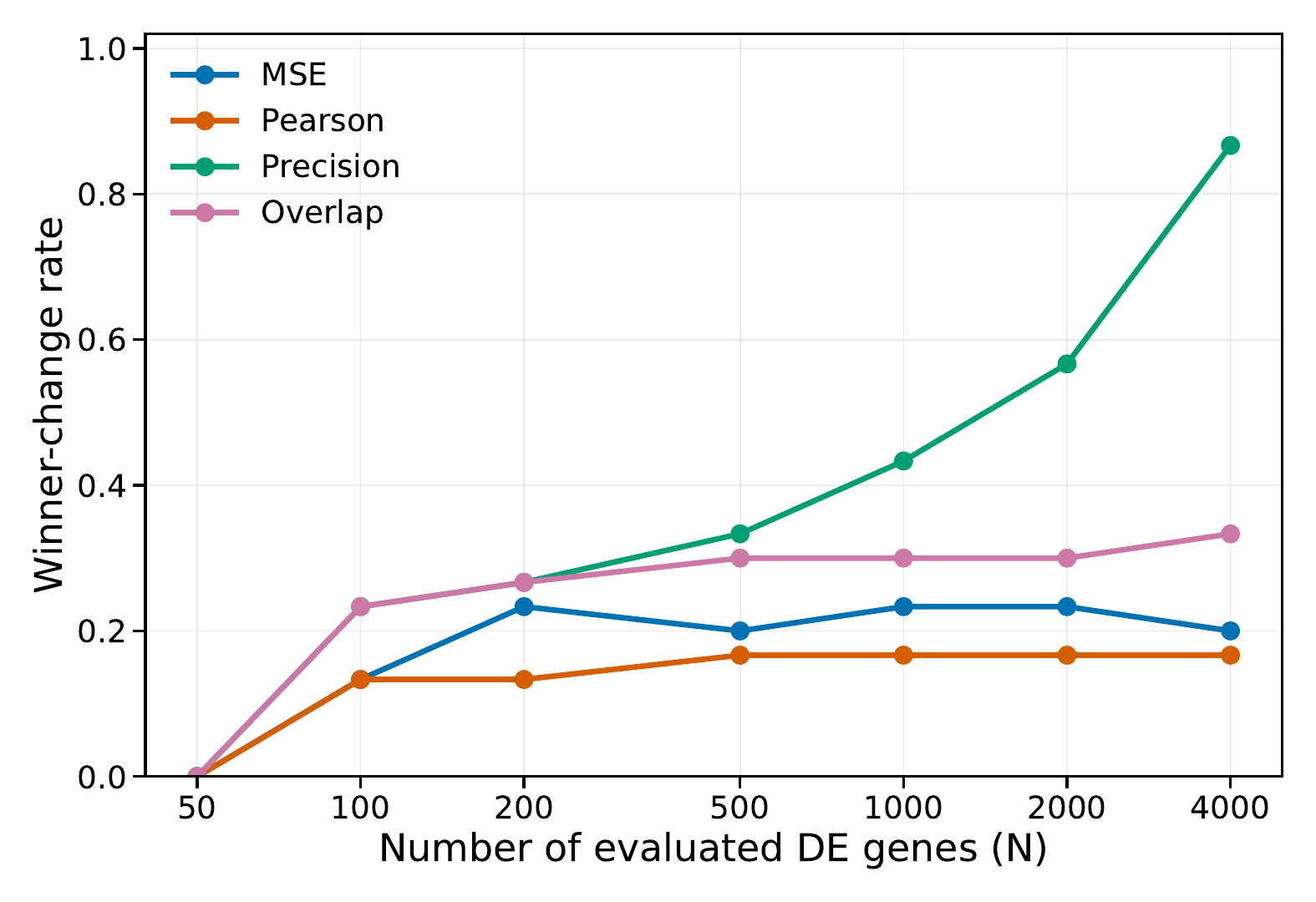}
            \caption{}
            \label{fig:winner_change}
        \end{subfigure}
    \end{minipage}

    \caption{
    \textbf{(a)} Model rankings as a function of the number \(N\) of
    evaluated DE genes. Each row corresponds to a dataset and each
    column to an evaluation metric. Rank \(1\) denotes the
    best-performing method. Different datasets have different number of DE genes.
    \textbf{(b)} Mean absolute rank displacement relative to the
    ordering obtained at \(N=50\), averaged across dataset--fold blocks.
    \textbf{(c)} Frequency with which the top-ranked method changes
    relative to the ordering obtained at \(N=50\).
    The final cutoff corresponds to \(4000\) genes or the maximum
    available number of eligible DE genes when fewer are present.
    }
    \label{fig:ranking_sensitivity}
\end{figure}

\subsection{Inflated predicted significance distorts Precision-based evaluation}
\label{sec:gene_set}

\subsubsection{Precision-based rankings change with {\it N}}
\label{sec:precision_rankings}

Figure~\ref{fig:rank_cross} shows that the evaluated gene-set size can
alter the ordering of competing methods. Rankings based on numerical
accuracy are comparatively stable after the smallest values of
\(N\), whereas Precision exhibits frequent rank crossings across datasets.
Overlap presents an intermediate pattern, with substantial changes in some
datasets but greater stability than Precision.
Mean absolute rank displacement measures how many ranking positions a method moves, on average, relative to its rank at \(N=50\), while the winner-change rate is the proportion of dataset--fold evaluations in which the top-ranked method differs from that at \(N=50\). Figures~\ref{fig:rank_displacement} and ~\ref{fig:winner_change} show that both quantities increase substantially for Precision as \(N\) grows, reaching an average displacement of \(2.58\) positions and a winner-change rate of \(86.7\%\) at the largest cutoff. In contrast, the rankings induced by MSE, Pearson, and Overlap remain substantially more stable, with mean displacements below one position and winner-change rates no greater than \(33.3\%\).
To test whether changing $N$ systematically alters model rankings, we compare every pair \((N_1,N_2)\) separately for each metric. After averaging model ranks across the five folds
within each dataset, we define the model-wise rank change as
\[
\Delta_{d,m}^{(k;N_1,N_2)}
=
\bar r_{d,m}^{(k,N_2)}
-
\bar r_{d,m}^{(k,N_1)},
\]
where \(k\) denotes the metric, \(d\) the dataset, and \(m\) the model.
We summarize systematic reordering as
\[
T^{(k;N_1,N_2)}
=
\frac{1}{M}\sum_{m=1}^{M}
\left(
\frac{1}{D}\sum_{d=1}^{D}
\Delta_{d,m}^{(k;N_1,N_2)}
\right)^2.
\]

For each metric and \(N\)-pair, we obtain an exact \(p\)-value by jointly
sign-flipping the complete rank-change vector of each dataset over all
\(2^D\) possible assignments. The resulting \(p\)-values are adjusted
across the 15 pairwise \(N\)-comparisons within each metric using the
Benjamini--Hochberg procedure.
Precision was the only metric for which systematic ranking changes
remained significant after correction. Twelve of the 15 comparisons
were significant
(\(p=0.03125\), \(q_{\mathrm{BH}}=0.03906\)), while the rest 
are still close to significance with (\(p=q_{\mathrm{BH}}=0.0625\)). This instability is also reflected in the relationship between detection metrics. Agreement between Precision and Overlap falls from \(\tau_b=0.83\) at \(N=50\) to \(0.03\) at \(N=4000\), while agreement between MSE and Pearson decreases only from \(0.77\) to \(0.66\). This can instill bias in case of scores that aggregate through all four metrics. We have a devoted section on this metric disagreement in App. \ref{sec:appendix_agreement}.

\subsubsection{Inflated predicted significance contributes to Precision instability}
\label{sec:precision_instability}

To understand why Precision becomes unstable while the rest remain comparatively robust, we will examine its definition and how the predicted significant-DE lists evolve with \(N\).
As defined in Sec.~\ref{sec:metrics}, Precision uses \(K_P=\min(N,|P|)\), where \(P\) is the list of predicted significant genes. This changes based on the model, in contrast to
Overlap, where it is determined by the observed significant-DE list and is therefore shared by all models for the same perturbation. In our evaluation, the real significant genes \(|R|\geq N\) account for 35\% of the overall cases, 
whereas the predicted significant genes \(|P|\geq N\) are approximately \(95\%\). 
The real significant gene set size $|R|$ starts at $70\%$ for $N=50$ and drops to $21\%$ for $N=1000$ and $6\%$ for $N=4000$ while the predicted $|P|$ go from $97\%$ to $95\%$ and $94\%$ respectively.
Consequently, in our experiments, while Overlap saturates due to the finite size of the observed DE set, Precision continues to evaluate longer prefixes because the predicted significant set is typically much larger than the observed one.

This excess of predicted significant genes does not necessarily imply that the predicted effects are larger. In fact, the results indicate that predicted significance can be inflated by an evaluation artifact. To be specific, of the \(620.9\) million tested perturbation--gene pairs, \(416.5\) million (\(67.1\%\)) passed the predicted \(FDR<0.05\) threshold. The pattern was consistent across methods, and even the AVG Ctrl baseline, which contains no perturbation-specific information, classified \(69.2\%\) of pairs as significant. However, \(36.3\) million significant calls (\(8.7\%\)) had a minuscule effect, i.e., \(|\log_2\mathrm{FC}|<0.1\), including some effects arbitrarily close to zero. These results show that predicted significance is not determined by effect magnitude alone.


We hypothesize that replication of perturbation-level predictions contributes to this excess significance. During evaluation, each predicted expression profile is repeated to match the observed cell count, adding no independent information while increasing the nominal sample size supplied to the MWU test. We next analyze how this mismatch affects the test statistic and nominal \(p\)-values, and explain how an enlarged predicted significant-DE set can contribute to the observed sensitivity of Precision to \(N\).

\subsubsection{Theoretical analysis on the Precision sensitivity}


To analyze the impact of replication of predicted expression profiles on the Precision evaluation, we revisit the mathematics underlying the Mann--Whitney \(U\) test and how it affects the DE gene lists.
Recall that $C, C_{\mathrm{ctrl}}$ are the sets of perturbed and non-targeting control cells, respectively.
For each gene, the statistics $U^\mathrm{obs}_g, U^\mathrm{pred}_g$ on observed and predicted gene expression are defined as:
\begin{equation}
\begin{aligned}
    U^\mathrm{obs}_g &= \sum_{i\in C} \sum_{j\in C_{\mathrm{ctrl}}} \left[ \mathbbm{1}\{X_{ig}>X^{\mathrm{ctrl}}_{jg}\} + \frac{1}{2} \mathbbm{1}\{X_{ig}=X^{\mathrm{ctrl}}_{jg} \} \right], \\
    U^\mathrm{pred}_g &= \sum_{i\in C} \sum_{j\in C_{\mathrm{ctrl}}} \left[ \mathbbm{1}\{\widehat{X}_{ig}>X^{\mathrm{ctrl}}_{jg}\} + \frac{1}{2} \mathbbm{1}\{\widehat{X}_{ig}=X^{\mathrm{ctrl}}_{jg} \} \right].
\end{aligned}
\end{equation}
To simplify the discussion, we denote $s\in\{\mathrm{obs}, \mathrm{pred}\}$ and omit the continuity correction.
For each gene, assume that ties occur only among the replicated predictions.
Let $t^s_g$ denote the size of this single tie group, with $t^\mathrm{obs}_g=0$ for the observed comparison, which has no ties.
With large samples, the standardized value $Z^s_g \xrightarrow{d} N(0,1)$, where:
\begin{equation}
\begin{aligned}
    Z^s_g &= \frac{U^s_g-\mu^s_g}{\sigma^s_g}, \qquad \mu_g^s=\frac{|C|\times|C_{\mathrm{ctrl}}|}{2}, \\
    (\sigma_g^s)^2 &= \frac{|C|\times|C_{\mathrm{ctrl}}|}{12}\Biggl[|C|+|C_{\mathrm{ctrl}}|+1-\frac{(t^s_g)^3-t^s_g}{(|C|+|C_{\mathrm{ctrl}}|)(|C|+|C_{\mathrm{ctrl}}|-1)}\Biggr].
\end{aligned}
\end{equation}
The two-sided $p$-values are:
\[
\begin{aligned}
p^\mathrm{obs}_g &= \mathrm{MWU}\left( \left( (X_{ig})_{i\in C}, (X_{jg}^{\mathrm{ctrl}})_{j\in C_{\mathrm{ctrl}}} \right) \right) = 2\left( 1-\Phi(|Z_g^\mathrm{obs}|)\right), \\
p^\mathrm{pred}_g &= \mathrm{MWU}\left( \left( (\widehat{X}_{ig})_{i\in C}, (X_{jg}^{\mathrm{ctrl}})_{j\in C_{\mathrm{ctrl}}} \right) \right) = 2\left( 1-\Phi(|Z_g^\mathrm{pred}|)\right),
\end{aligned}
\]
where $\Phi$ is the cumulative distribution function of the standard normal distribution.

Deterministic models and models that ignore cell heterogeneity can only generate a single prediction, because their prediction is purely based on the perturbation condition and target gene, without considering the variability among individual cells' gene expression profiles.
During MWU test, $Z^s_g$ is asymptotically normal because we assume that the sample sizes $|C|$ and $|C_{\mathrm{ctrl}}|$ are sufficiently large.
For models with single predictions, the prediction group contains only one profile and the assumption is violated.

Replicating a prediction can increase the nominal sample size. 
However, this replication strategy can also increase the number of predicted significant DE genes.
Suppose we generate $c$ replicates for each prediction, then $|C|=c$ and $\widehat{X}$ has $c$ replicated rows.
The single tie group then has size $t^\mathrm{pred}_g=c$. Define
\begin{equation}
    B(c,|C_{\mathrm{ctrl}}|) = c+|C_{\mathrm{ctrl}}|+1-\frac{c^3-c}{(c+|C_{\mathrm{ctrl}}|)(c+|C_{\mathrm{ctrl}}|-1)}.
\end{equation}
Writing $U^\mathrm{pred}_g(c)$ and $Z^\mathrm{pred}_g(c)$ for the statistics with $c$ copies, the tie-corrected variance used by the test is $(\sigma_g^\mathrm{pred}(c))^2=c|C_{\mathrm{ctrl}}|B(c,|C_{\mathrm{ctrl}}|)/12$. Thus,
\begin{equation}
    U^\mathrm{pred}_g(c) = c\,U^\mathrm{pred}_g(1),\quad
    Z^\mathrm{pred}_g(c) = Z^\mathrm{pred}_g(1)\sqrt{\frac{c(|C_{\mathrm{ctrl}}|+2)}{B(c,|C_{\mathrm{ctrl}}|)}}.
\end{equation}
For $c=|C_{\mathrm{ctrl}}|=100$, the inflation factor for $Z^\mathrm{pred}_g$ relative to the single-prediction standardized statistic is approximately $7.62$.
As $c$ increases, the inflation factor for $Z^\mathrm{pred}_g$ also increases, leading to an overestimation of statistical significance.
The observed perturbed $X$ remains unchanged with $|C|$ original samples, whereas the prediction side is inflated by the replication strategy.
Thus, the predicted DE gene list $P$ is longer, $|P|$ is larger, and the inflation propagates to $K_P$ and the calculation of Precision.

Although the significance-estimation artifact is present at every cutoff $N$, smaller cutoffs yield comparatively more stable rankings and stronger agreement between Precision and Overlap. Since the predicted significant-DE list is ordered by absolute predicted log-fold change, smaller cutoffs emphasize the strongest predicted effects which have a higher chance of really being significant. 
Moreover, Precision and Overlap are identical whenever both the observed and predicted significant-DE lists contain at least $N$ genes, a condition that is more readily satisfied at smaller cutoffs. As $N$ increases beyond the observed significant-DE count, Precision can continue evaluating longer predicted prefixes while Overlap stops expanding, exposing the asymmetry between the two lists, as indicated also by the percentages in Sec. \ref{sec:precision_instability}. 


\subsection{Benchmark}

For the benchmark comparison, we restrict the evaluation to $N \in \{50,100,200\}$ as a pragmatic choice based on the observed ranking stability. This restriction does not correct the underlying significance estimates but limits evaluation to the strongest retained effects and hence reduces the influence of the artifact on Precision. We aggregate scores separately for each method and metric in
three stages. First, within each dataset, we compute a mean score for each cutoff by averaging over the folds in which that cutoff was evaluated. Missing scores are excluded, rather than replaced by scores from smaller cutoffs. Second, we average these cutoff-specific means with equal weight to obtain one dataset-level score, excluding cutoffs with no available results. Thus, cutoffs evaluated in more folds do not receive greater weight in the dataset-level average. Finally, we average the dataset-level scores across datasets with available results, giving each dataset equal
weight.


Overall the benchmark shows that greater architectural complexity does not consistently translate into better performance.  RF-GO achieves the highest Pearson, Precision, and Overlap, while RF-scf obtains the lowest MSE and ranks second on the remaining
metrics (Table~\ref{tab:model_performance}). Notably, AVG Train ranks third in MSE, Pearson, and Overlap, demonstrating that even a simple averaging baseline remains competitive as indicated in the literature. These findings agree with part of the literature on simpler methods \citep{RfgoPaper,vinas2025systema}; however, they disagree that simple regression models such as Bilinear are highly competitive \citep{ahlmann2025deep}. We hypothesize that this is because that work's conclusions rely partly on top-1000 expressed, not DE, genes, which is a very specific evaluation not used extensively in the literature. Finally, more complex models such as GEARS and LPM also underperformed compared to RF.  Results disaggregated by dataset and cutoff are provided in App.~\ref{sec:heatmaps}.

\begin{table}[!ht]
    \centering
    \caption{Model performance across evaluation metrics. All values $\times 100$.}
    \label{tab:model_performance}
        \begin{tabular}{lrrrr}
            \toprule
             & MSE & Pearson & Precision & Overlap \\
            \midrule
            Bilinear      & 14.31 & 47.70 & 6.15 & 6.09 \\
            LPM           & 25.14 & 26.70 & 4.21 & 4.42 \\
            Ridge-scf     & \textsc{12.32} & 45.96 & 6.06 & 6.37 \\
            GEARS         & 16.53 & 45.21 & 6.29 & 6.33 \\
            AVG Ctrl       & 14.59 & 33.23 & 4.15 & 4.69 \\
            AVG Train       & 12.40 & 48.92 & 6.15 & 6.65 \\
            RF-scf        & \textbf{11.77} & \textsc{51.92} & \textsc{6.61} & \textsc{7.10} \\
            RF-GO         & 12.68 & \textbf{54.77} & \textbf{7.20} & \textbf{7.19} \\
            \bottomrule
        \end{tabular}%
\end{table}


\section{Conclusion}
We benchmarked eight methods and baselines across six single-gene perturbation datasets in order to discern the role of different cutoffs for DE genes in the results of current methods. We found that Precision changes significantly as $N$ increases and we identified an evaluation artifact that contributes to this instability. Repeating predictions across cell-level rows to match the ground truth expression matrix can inflate predicted DE significance, which leads to admitting weak effects into the predicted significant-DE lists. Since Precision determines its effective cutoff from the predicted rather than observed significant-DE count, this effect increasingly influences evaluation as $N$ grows. 
Restricting the benchmark comparison to smaller cutoffs does not alleviate the effect, but since the majority of genes at these cutoffs are significant, it diminishes it. To this end, we take an aggregate of the results for $N=\{50,100,200\}$. Under this regime, random forests using gene ontology features or scFoundation representations achieve the strongest results, while a training-mean baseline remains competitive.

Future work will develop and validate evaluation procedures suited to perturbation-level point predictions, rather than treating duplicated profiles as independent cellular observations. For small numbers of independent cell-level predictions, particularly for deterministic models, the exact MWU test should be used, so as to avoid artificially replicating predictions to increase the nominal sample size. Finally, we plan to extend the benchmark to other advanced methods \citep{cui2024scgpt,adduri2025predicting,txpert}, and add finetuning versions of the large models.

\section*{Statements}

\subsection*{AI use statement}
We used AI to identify assist with literature searches, typographical errors and improve language and LaTeX formatting. 
The authors reviewed the material incorporated into the manuscript and take full responsibility for its content.

\subsection*{Reproducibility statement}
Data preprocessing is described in
Appendix~\ref{sec:preprocessing}. Section~3 describes the evaluation
protocol and hyperparameter settings, while
Section~\ref{sec:precision_rankings} details the statistical tests
for ranking changes. The supplementary code includes a README
with instructions for reproducing the results and documentation
of model-specific data handling and implementation details.

\subsection*{Ethics statement}
This work analyzes existing single-cell perturbation datasets
to assess the reliability of model evaluation. Benchmark
performance should not be interpreted as evidence of therapeutic
efficacy, and downstream biological applications require
independent experimental validation.

\bibliography{iclr2027_conference}
\bibliographystyle{iclr2027_conference}

\appendix

\section{Appendices}

\subsection{Data processing}
\label{sec:preprocessing}
\begin{figure}[h!]
    \centering
    \includegraphics[width=\linewidth]{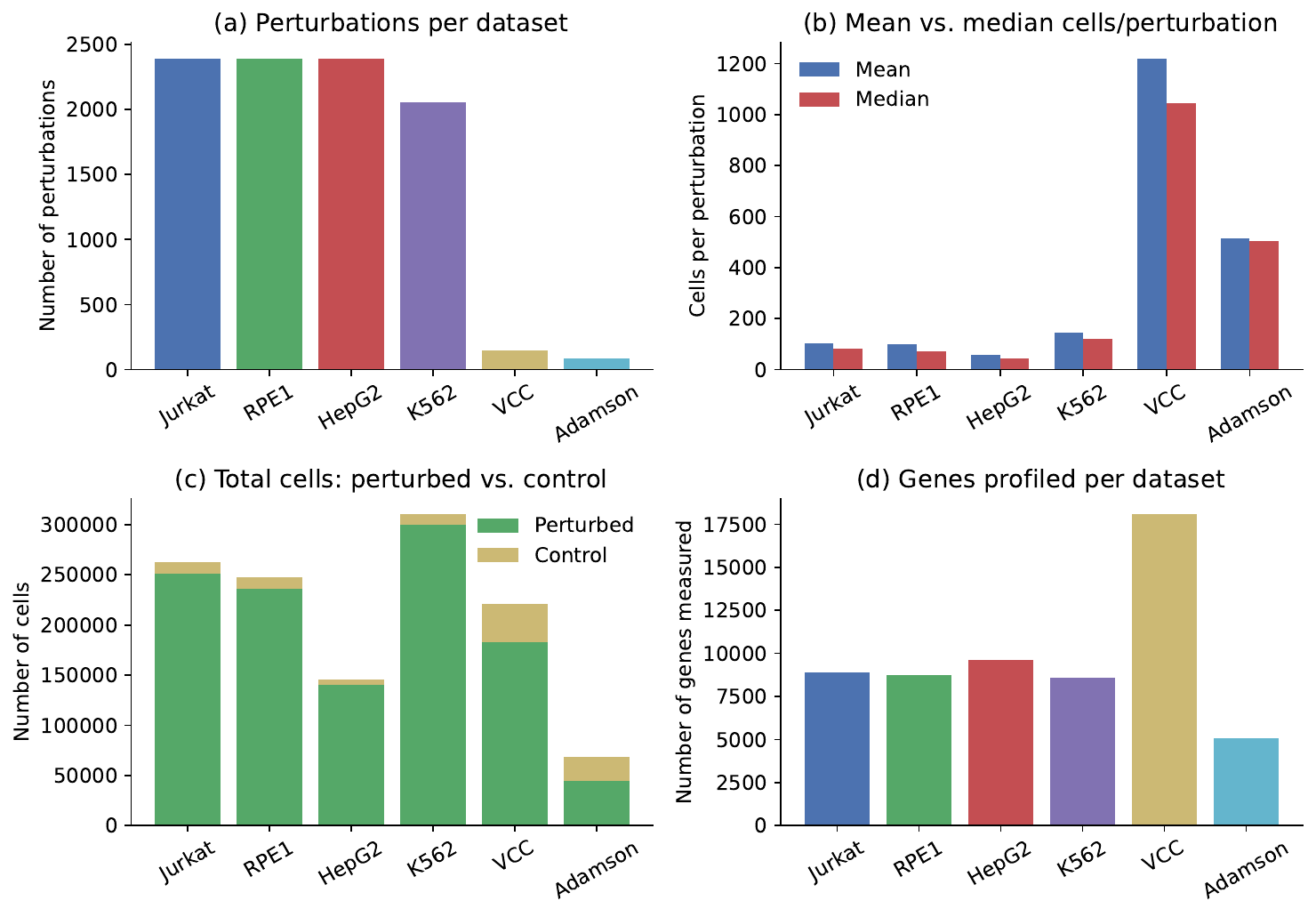}
    \caption{Overview of dataset characteristics. (a) Number of
    perturbations per dataset. (b) Mean vs.\ median cells per
    perturbation, illustrating the right-skewed coverage typical of
    single-cell perturbation screens. (c) Total cells per dataset,
    split into perturbed and non-targeting control cells. (d) Number
    of genes profiled per dataset.}
    \label{fig:dataset_overview}
\end{figure}

For each dataset we load the provided \texttt{AnnData} object and
harmonize it to a common schema. Gene identifiers are mapped to gene
symbols and a \texttt{target\_gene} field is added to every cell,
recording either the perturbed gene or the control label
\texttt{non-targeting}; control-cell labels from heterogeneous source
naming conventions are unified to this single label. Duplicated gene
names are removed, and any perturbation whose target gene is not
itself among the measured/profiled genes is dropped, since such
perturbations cannot be scored against a matching expression readout.
For the Adamson dataset specifically, the provided expression matrix
is supplied in natural-log1p-normalized form rather than raw UMI
counts; we invert this transform ($\exp(x)-1$) to recover
linear-normalized expression values, consistent with the scale used
by the other five datasets, while preserving matrix sparsity.
Non-targeting control cells are separated from perturbed cells and
held out as a dedicated control set, used both for baseline
predictors and as a reference profile during evaluation.
An overview of the dataset characteristics and comparison can be seen in Fig. \ref{fig:dataset_overview}.

\subsection{Methods}
\label{sec:app_methods}

In detail the methods we use are:





  \textbf{GEARS} \citep{roohani2024gears}: The neural network predicts an expression change and adds it to an input control cell’s expression. At inference, it uses sampled control cells and averages their predicted post-perturbation profiles to obtain one expression profile per perturbation.

  \textbf{Ridge/RF-scf} \citep{scf}: Contextual scFoundation embeddings describe the perturbed gene. A ridge regression or a random forest model learns the difference between mean perturbed expression and mean control expression. Its predicted change is added to the control mean to reconstruct post-perturbation expression.

  \textbf{RF-GO} \citep{RfgoPaper}: A random forest maps GO features of the perturbed gene directly to mean post-perturbation expression. Its training targets are expression profiles, so there is no explicit control-subtraction and control-addition step.

  \textbf{Bilinear} \citep{ahlmann2025deep}: The model learns expression changes relative to a baseline b, defined as the per-gene mean across training perturbation profiles. It fits the model to baseline-subtracted expression and reconstructs post-perturbation expression. 
  
  \textbf{LPM} \citep{lpm}: The Large Perturbation Model is trained on mean expression profiles grouped by perturbation and context. The benchmark obtains predicted expression for each perturbation–gene pair and assembles
  these values into a post-perturbation profile, without an explicit control-addition step in the wrapper. 

  \textbf{Average control} \citep{vinas2025systema}: Every perturbation receives the mean expression profile of all the non-targeting control cells (not batch based). This represents an assumption of no perturbation-induced change before output transformation.

  \textbf{Average train} \citep{vinas2025systema}: Every perturbation receives the mean expression profile across all perturbed training cells (not batch based). The prediction is identical for every test perturbation.


  Before evaluation, the wrappers clip negative predicted expression to zero and apply log1p. Methods producing one profile per perturbation repeat that profile across the required prediction rows.
For methods evaluation we use 5-fold cross-validation performed at the
\emph{perturbation} level rather than the cell level: unique target
genes are randomly partitioned into five folds, so that all cells
belonging to a given perturbation fall entirely within either the
training or the test split. This tests each model's ability to
generalize to genes it has not seen perturbed during training, rather
than merely interpolating between cells of an already-observed
perturbation. Depending on configuration, non-targeting control cells
are either excluded from both splits and reserved solely for
control-based baselines and evaluation, or retained in the training
split only, never appearing in the test set. All splits use a fixed
random seed for reproducibility.
Beyond this common pipeline, a subset of models require additional,
model-specific dataset adaptations applied at run time (e.g., GEARS
and LPM expect the control label \texttt{ctrl} rather than
\texttt{non-targeting}, among other format requirements); these
adaptations are not part of the shared preprocessing described above.


\section{Changes in Metric Agreement}
\label{sec:appendix_agreement}

\begin{figure}[h!]
    \centering
    \begin{subfigure}[c]{0.35\linewidth}
        \centering
        \includegraphics[width=\linewidth]{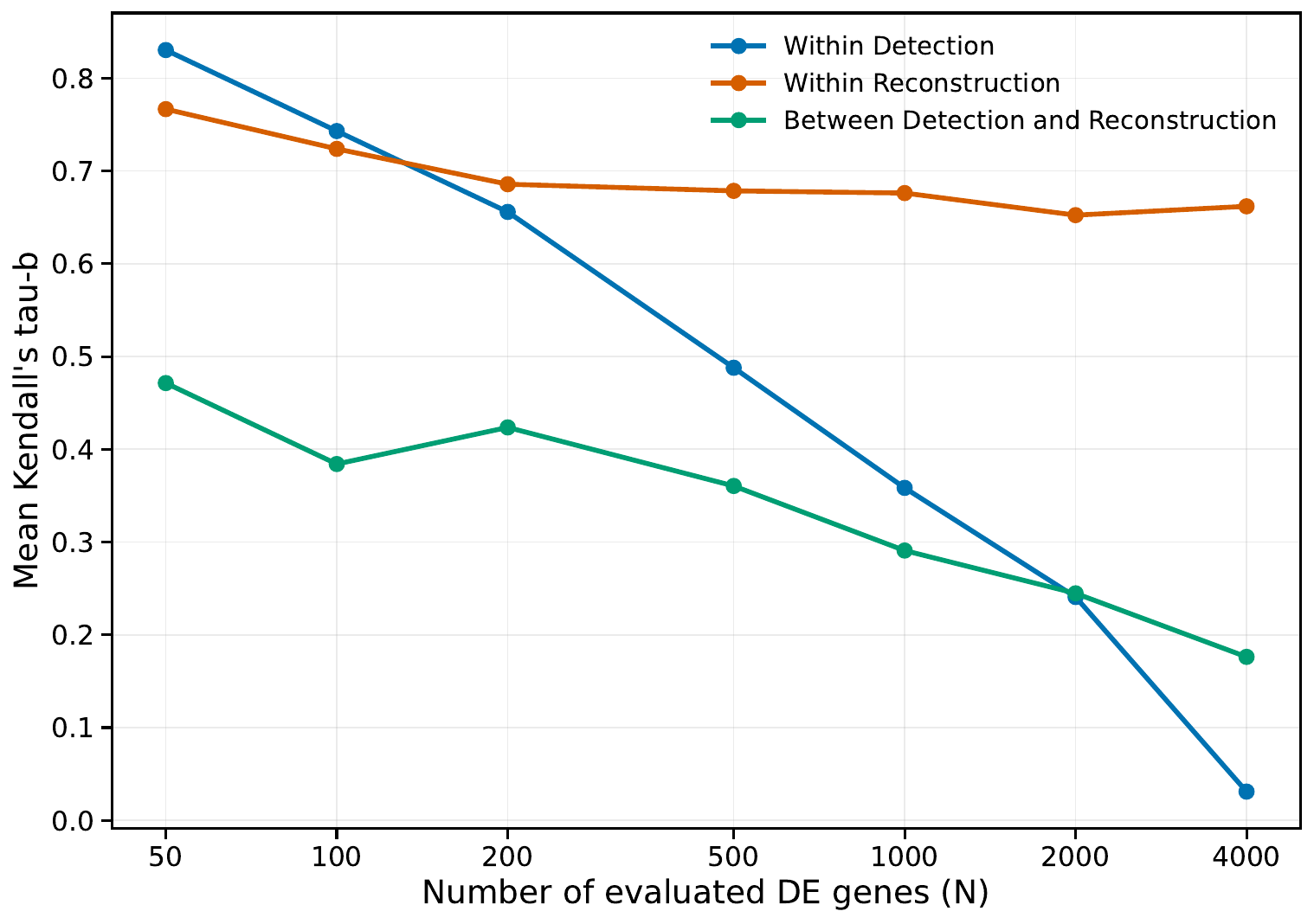}
        \caption{}
        \label{fig:agreement_by_n}
    \end{subfigure}%
    \hfill
    \begin{subfigure}[c]{0.65\linewidth}
        \centering
        \includegraphics[width=\linewidth]{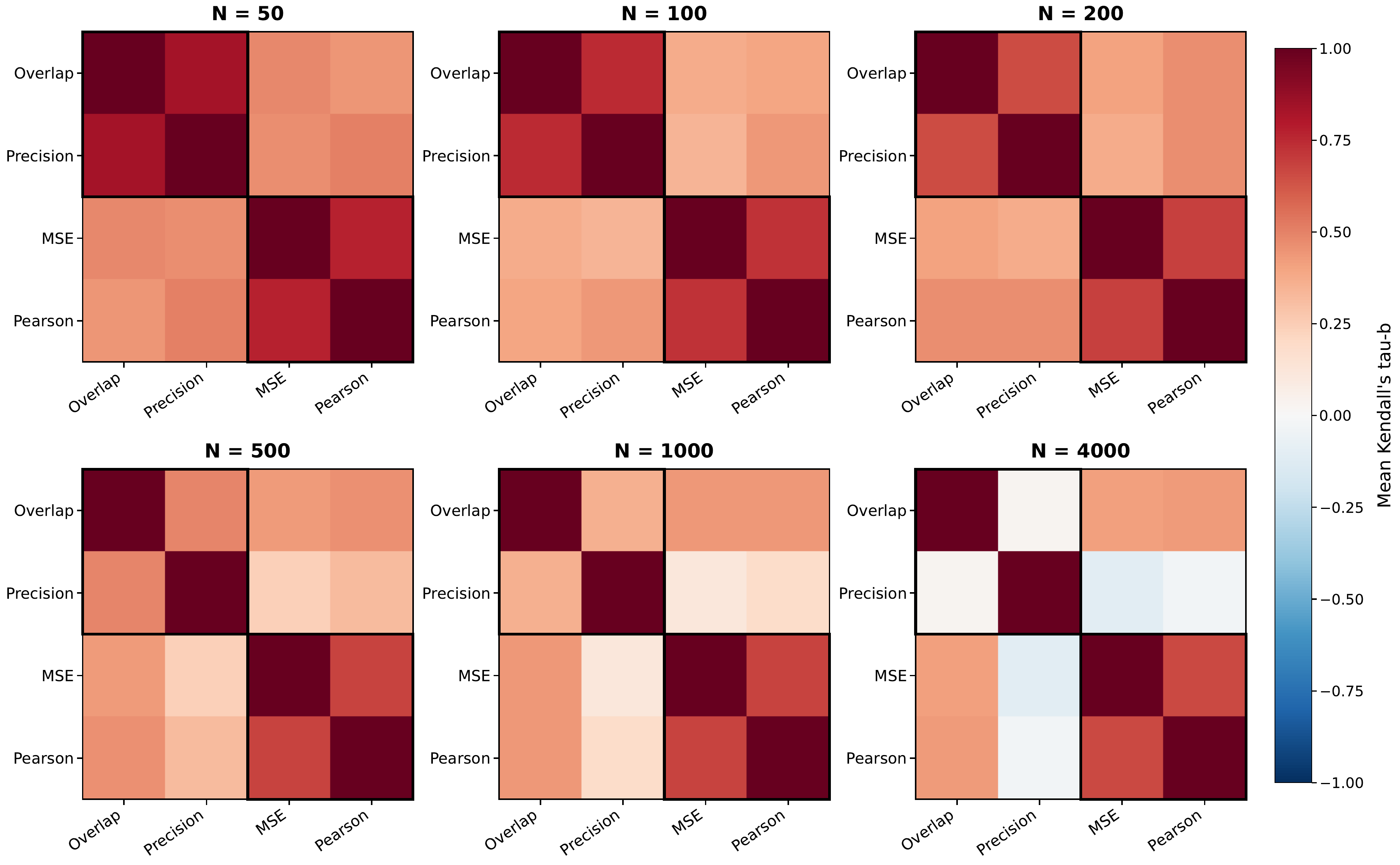}
        \caption{}
        \label{fig:metric_correlation_heatmaps}
    \end{subfigure}
    \caption{(a) Agreement among metric-induced model rankings as a function of the number \(N\) of evaluated DE genes (mean Kendall's \(\tau_b\), averaged across folds within each dataset and then across datasets with equal weight). Detection$=\{$Overlap, Precision$\}$, reconstruction$=\{$MSE and Pearson$\}$. (b) Pairwise Kendall's \(\tau_b\) between metric-induced model rankings at representative values of \(N\). Black rectangles indicate metric families. Values are averaged across folds within each dataset and then across datasets with equal weight.}
    \label{fig:agreement_and_heatmaps}
\end{figure}

Changes in \(N\) affect not only model scores and rankings, but also the agreement among the metrics. We therefore rank the methods independently under each metric within every dataset--fold block and compute Kendall's \(\tau_b\) between the resulting rankings. As shown in Figure~\ref{fig:agreement_by_n}, agreement between Overlap and Precision decreases monotonically from \(0.83\) at \(N=50\) to \(0.03\) at \(N=4000\), whereas agreement between MSE and Pearson remains comparatively stable, decreasing only from \(0.77\) to \(0.66\). Agreement between detection and reconstruction also weakens as \(N\) increases.
Figure~\ref{fig:metric_correlation_heatmaps} localizes this divergence to the detection metrics. In accordance with section \ref{sec:precision_rankings}, precision and overlap agree at $N=50$ but, due to significant changes in precision, they gradually discord.
Consequently, composite evaluation schemes that average heterogeneous metrics are sensitive to the choice of \(N\) and may instill bias towards one metric family over the other.

\section{Benchmark Heatmap}
\label{sec:heatmaps}

\begin{figure}[t]
    \centering
    \includegraphics[width=0.8\linewidth]{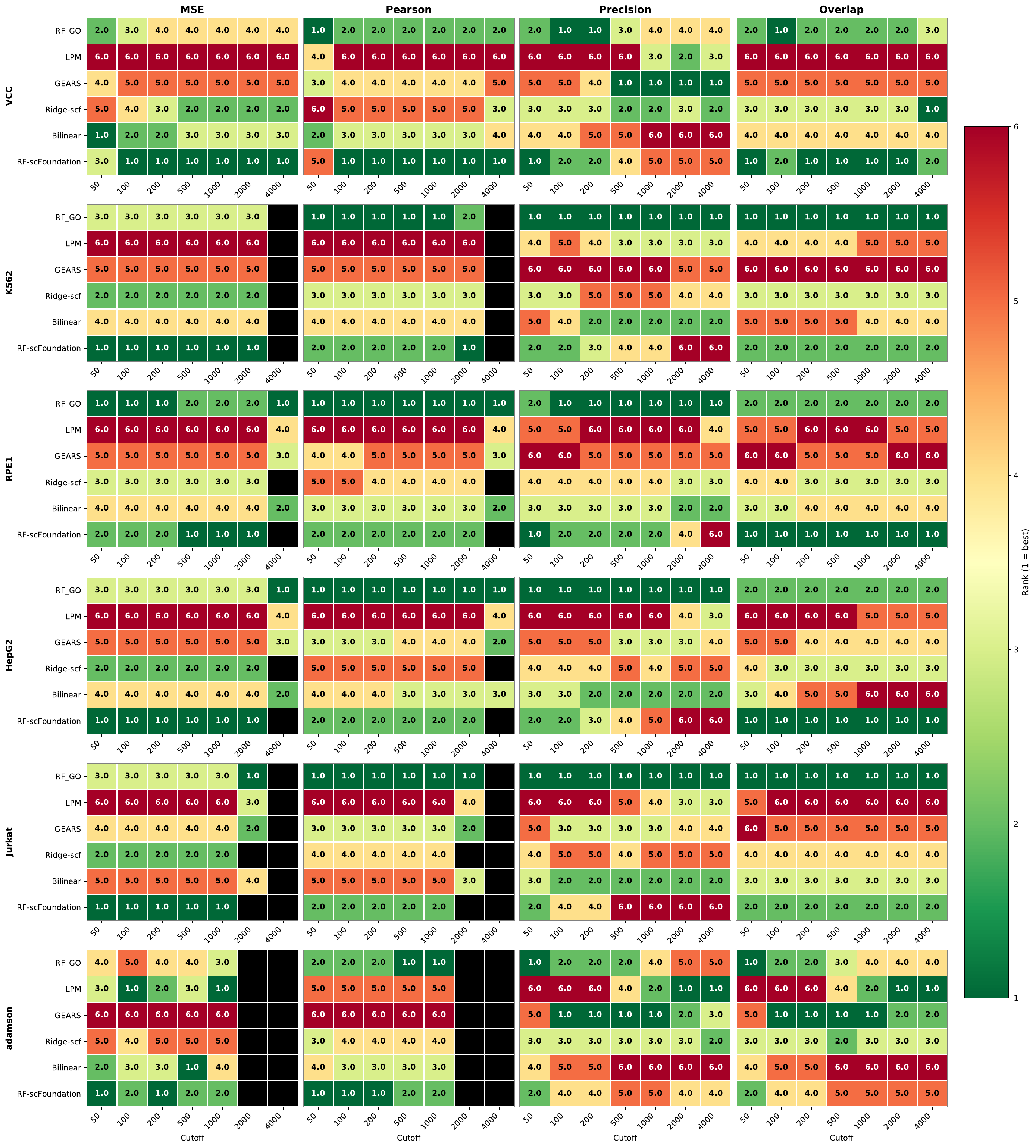}
    \caption{Heatmap showing the rank of each evaluated model. Lower ranks indicate better performance. Black cells indicate that no DE genes were available for the corresponding value of $N$.}
    \label{fig:overall_method_ranks}
\end{figure}

Fig.~\ref{fig:overall_method_ranks} summarizes the benchmark by showing, for every non-baseline model and evaluation metric, the rank achieved across the range of $N$ values considered.

\end{document}

%% file: math_commands.tex
\usepackage{amsmath,amsfonts,bm}

\def\eqref#1{equation~\ref{#1}}

\def\1{\bm{1}}

\def\ra{{\textnormal{a}}}

\def\rx{{\textnormal{x}}}

\def\rva{{\mathbf{a}}}

\def\erva{{\textnormal{a}}}

\def\ervx{{\textnormal{x}}}

\def\rmA{{\mathbf{A}}}

\def\vmu{{\bm{\mu}}}
\def\vtheta{{\bm{\theta}}}
\def\va{{\bm{a}}}

\def\ve{{\bm{e}}}

\def\vx{{\bm{x}}}

\def\eva{{a}}

\def\mA{{\bm{A}}}

\def\mH{{\bm{H}}}
\def\mI{{\bm{I}}}
\def\mJ{{\bm{J}}}

\def\mX{{\bm{X}}}

\def\mSigma{{\bm{\Sigma}}}

\DeclareMathAlphabet{\mathsfit}{\encodingdefault}{\sfdefault}{m}{sl}
\SetMathAlphabet{\mathsfit}{bold}{\encodingdefault}{\sfdefault}{bx}{n}
\newcommand{\tens}[1]{\bm{\mathsfit{#1}}}
\def\tA{{\tens{A}}}

\def\tX{{\tens{X}}}

\def\gG{{\mathcal{G}}}

\def\sA{{\mathbb{A}}}
\def\sB{{\mathbb{B}}}

\def\sS{{\mathbb{S}}}

\def\emA{{A}}

\newcommand{\etens}[1]{\mathsfit{#1}}

\def\etA{{\etens{A}}}

\newcommand{\E}{\mathbb{E}}

\newcommand{\R}{\mathbb{R}}

\newcommand{\KL}{D_{\mathrm{KL}}}
\newcommand{\Var}{\mathrm{Var}}

\newcommand{\Cov}{\mathrm{Cov}}

\newcommand{\normltwo}{L^2}
\newcommand{\normlp}{L^p}

\newcommand{\parents}{Pa} 